\documentclass[amsmath,jcp,amssymb,aip,graphicx,reprint]{revtex4-1}
\usepackage{graphicx}
\usepackage{etoolbox}
\usepackage{multirow}

\usepackage{color}
\RequirePackage{color}

\definecolor{Myorange}{cmyk}{0,0.42,1,0}

\newcounter{subeqn} %

\draft 

\newcommand{\wfae}{| \psi \rangle}
\newcommand{\wfps}{| \tilde{\psi} \rangle}
\newcommand{\pawT}{\mathcal{T}}

\newcommand{\js}{$^1$J}
\newcommand{\jd}{$^2$J}

\newcommand{\bmu}{\boldsymbol\mu}

\newcommand{\ket}    [1]{{|#1\rangle}}
\newcommand{\bra}    [1]{{\langle#1|}}

\newcommand{\kunits}{$10^{19}\thinspace\cdot\thinspace$T$^2\thinspace\cdot\thinspace$J$^{-1}$}

\renewcommand\Re{\operatorname{Re}}

\begin{document}

\title{Relativistic nuclear magnetic resonance J-coupling with ultrasoft pseudopotentials and the zeroth-order regular approximation}

\author{Timothy F. G. Green}
 \email{tim.green@materials.ox.ac.uk}
\author{Jonathan R. Yates}%
 \email{jonathan.yates@materials.ox.ac.uk}
\affiliation{ 
  Department of Materials, University of Oxford, UK
}

\date{\today}

\begin{abstract}
We present a method for the first-principles calculation of nuclear magnetic resonance (NMR) J-coupling in extended systems using state-of-the-art ultrasoft pseudopotentials and including scalar-relativistic effects. The use of ultrasoft pseudopotentials is allowed by extending the projector augmented wave (PAW) method of Joyce et. al [J. Chem. Phys. 127, 204107 (2007)]. We benchmark it against existing local-orbital quantum chemical calculations and experiments for small molecules containing light elements, with good agreement. Scalar-relativistic effects are included at the zeroth-order regular approximation (ZORA) level of theory and benchmarked against existing local-orbital quantum chemical calculations and experiments for a number of small molecules containing the heavy row six elements W, Pt, Hg, Tl, and Pb, with good agreement. Finally, \js(P-Ag) and \jd(P-Ag-P) couplings are calculated in some larger molecular crystals and compared against solid-state NMR experiments. Some remarks are also made as to improving the numerical stability of dipole perturbations using PAW.
\end{abstract}


\maketitle

\vskip2pc
\marginparwidth 3.1in
\marginparsep 0.5in

\section{\label{sec:introduction}Introduction}

Modern high resolution solid-state NMR experiments \cite{lesage_recent_2009,ashbrook_recent_2009} are a valuable tool for materials characterisation due to their sensitivity to the local atomic environment. Importantly, solid-state NMR can provide information on materials with compositional, positional or dynamic disorder\cite{florian_beyond_2013}. However, there is no straight-forward analytic technique to obtain atomic-level structure directly from an NMR spectrum; a `Bragg's law' for NMR. Instead, one must pursue computational-theoretical prediction of the NMR parameters that influence a spectrum in order to fully take advantage of the information present to interpret and assign spectra. First principles predictions of NMR parameters can also assist in the design of NMR experiments, such as determining observability and orientation of tensors. Overall, first-principles calculations offer the ability to fully exploit the information in experimental NMR data.

The sensitivity of NMR experiments to molecular geometry and electronic structure is a `double-edged sword,' being both an important chemical probe and a challenge to the computational theorist. For small, finite, systems, NMR parameters such as magnetic shielding, electric field gradients and J-coupling can be routinely calculated with quantum chemical methods based on local orbitals and have demonstrated value in assigning solution-state spectra \cite{helgaker_recent_2012}. Treatment of solid-state NMR systems with these methods requires the creation of finite-clusters, which need careful convergence with respect to the size of the cluster to ensure that the appropriate electronic environment is reconstructed. They also require careful selection of the basis set used to represent the wave function to ensure numerical convergence. A planewave approach with pseudopotentials is appealing for its algorithmic efficiency, automatic inclusion of periodic boundary conditions and easy systematic convergence of basis sets via the maximum kinetic energy of the waves used. However, since such calculations require the use of pseudopotentials, the calculated pseudo-wave function is non-physical near the nucleus, the very region that is so influential to NMR parameters. The development of the gauge-including projector augmented wave (GIPAW) method\cite{pickard_all-electron_2001} has enabled calculations of magnetic shielding in extended systems using pseudopotentials by reconstructing the form of the all-electron wavefunction near the nucleus. Extensive reviews are available in Refs. \onlinecite{bonhomme_first-principles_2012} and \onlinecite{charpentier_paw/gipaw_2011}.

This paper concerns itself with the theoretical prediction of NMR J-coupling, or indirect spin-spin coupling, particularly in solid-state systems with heavy ions. J-coupling is the indirect magnetic coupling between two nuclei mediated via the bonding electrons. It manifests in NMR spectra as fine structure splitting of resonant peaks, providing information on bonds such as strength, angles and the connectivity network. J-coupling has been well-studied in the gas and solution state for many decades, as the multiplet splitting in peaks is well resolved due to molecular tumbling decoupling anisotropic interactions. In contrast, solid-state J-coupling studies are more challenging due to anisotropic broadening. Recent advances in solid-state NMR experiments, such as higher MAS spinning rates (up to 90KHz) and ultra-high magnetic field strengths (up to 23.3T), have resulted in increased experimental and theoretical interest\cite{massiot_detection_2010} in measurements of J.

Joyce et al.\cite{joyce_first_2007} developed a method to calculate J-coupling constants from first-principles in extended systems within a planewave-pseudopotential density-functional theory (DFT) framework, using PAW to reconstruct the all-electron properties of the system. This method has been validated for a small number of systems containing light atoms against quantum chemical calculations and against experimental data \cite{joyce_density_2008,griffin_high-resolution_2010,bonhomme_new_2010,hung_probing_2011,florian_elucidation_2012}. There is great interest in making this a `full periodic table' method, i.e. being able to reliably treat systems containing any elements. However, it is known\cite{pyykko_relativistic_1988,enevoldsen_relativistic_2000} that J-coupling in systems containing heavy ions is extremely sensitive to the effects of special relativity. This is because both core states and valence states near the nucleus attain high kinetic energy and so should be treated using the Dirac equation, leading to contraction in the wave function and corrections to the operators representing electromagnetic (EM) interactions. Scalar relativistic effects (i.e. ignoring spin-orbit terms) in particular have been found to be the dominant correction in full four-component Dirac equation calculations\cite{enevoldsen_relativistic_2000}, at least for one-bond couplings. Autschbach and Ziegler\cite{autschbach_nuclear_2000,autschbach_nuclear2_2000} developed the application of the zeroth-order regular approximation (ZORA), an approximation to the Dirac equation, using DFT to the prediction of J-coupling in an all-electron, local orbital framework. Yates et al.\cite{yates_relativistic_2003} developed the use of ZORA with pseudopotentials and PAW for the calculation of NMR chemical shifts in systems containing heavy ions. In this paper we will incorporate the scalar-relativistic terms of Autschbach and Ziegler's ZORA approach within a planewave pseudopotential DFT\cite{RMP-Payne} framework to give a highly efficient method for predicting J-coupling within extended systems containing heavy ions at negligible extra computational cost as compared to the non-relativistic method. We also provide some improvements to the non-relativistic method of Joyce et al.\cite{joyce_first_2007}, removing some of the numerical difficulties present in that approach, and generalising the method to use state-of-the-art ultrasoft pseudopotentials\cite{vanderbilt_soft_1990}. We carefully benchmark our planewave-pseudopotential implementation against both experiment and existing quantum chemical calculations, and conclude by examining the effects of relativity on J-couplings in two Ag-containing molecular crystals.

\section{\label{sec:theory}Theory}

\subsection{\label{sec:theory-intro}Introduction}

We will proceed by first reviewing the derivation of the zeroth-order regular approximation from the Dirac equation. Then we will show how a scalar-relativistic theory of NMR J-coupling can be derived from the ZORA Hamiltonian. We then discuss Bl\"ochl's PAW as a general formalism for the calculation of all-electron properties from pseudopotential calculations and derive a form of the scalar-relativistic theory that is suitable for efficient pseudopotentials calculations.

\subsection{\label{sec:theory-zora}Zeroth-order regular approximation}

We start with the time-independent single particle Dirac equation\cite{dirac_quantum_1928} for an electron in Hartree atomic units,

\begin{equation}
[c\mathbf{\alpha}\cdot\hat{\mathbf{p}} + \beta c^2 + V]\psi = E \psi,
\end{equation}

where $\alpha$ and $\beta$ are the Dirac matrices and in the case of density-functional theory\cite{DFT-KS} $V$ represents the nuclear, Hartree and exchange-correlation potentials. The wave function $\psi$ is a complex four-component spinor, alternatively expressed in terms of the small component $\chi$ and the large component $\phi$. We can eliminate the small component $\chi$ by substitution,

\begin{equation}
  \chi = \hat{X} \phi = \frac{c \mathbf{\sigma}\cdot{\hat{\mathbf{p}}}}{2c^2 + E - V} \phi,
\end{equation}

and retrieve an energy-dependent Hamiltonian in the large component $\phi$ only:

\begin{equation}
  \hat{H}^{\textit{esc}}\phi = E\phi = V\phi + \frac{1}{2}\mathbf{\sigma}\cdot{\hat{\mathbf{p}}} (1 - \frac{E-V}{2c^2})^{-1} \mathbf{\sigma}\cdot{\hat{\mathbf{p}}}.
\end{equation}

To normalize $\phi$ we introduce a normalization operator $\hat{O} = \sqrt{1 + \hat{X}^\dagger\hat{X}}$ and so we find the transformed Hamiltonian:

\begin{align}
  \hat{H} &= (1 + \hat{X}^\dagger\hat{X})^{\frac{1}{2}} [V + c\mathbf{\sigma}\cdot{\hat{\mathbf{p}}} \hat{X}] (1 + \hat{X}^\dagger\hat{X})^{-\frac{1}{2}}
\end{align}

The standard expansion of $\hat{X}$ and $\hat{H}^{\textit{esc}}$ in $(E-V)/(2c^2)$ to give the relativistic Pauli approximation is appropriate when the classical velocity of the electrons is small compared to the speed of light. This breaks down for a nuclear Coulomb potential. Instead, following van Lenthe \cite{lenthe_relativistic_1993,lenthe_phd}, we expand in $1/(2c^2 - V)$, which is justified even near the singularity of a nuclear Coulomb potential. This gives

\begin{widetext}
\begin{equation}
  \hat{H}^{\textit{esc}} \approx V + \mathbf{\sigma}\cdot{\hat{\mathbf{p}}} \frac{c^2}{2c^2 - V} \mathbf{\sigma}\cdot{\hat{\mathbf{p}}} - \mathbf{\sigma}\cdot{\hat{\mathbf{p}}} \frac{c^2}{2c^2 - V} \frac{E}{2c^2 - V} \mathbf{\sigma}\cdot{\hat{\mathbf{p}}} + ... .
  \label{eqn:ra-expansion}
\end{equation}
\end{widetext}

To lowest order the expansion of $\hat{O}$ in $1/(2c^2 - V)$ gives the identity, so we find that the first two terms of Eqn.~\ref{eqn:ra-expansion} are the ZORA Hamiltonian:

\begin{equation}
  \hat{H}^{\textit{ZORA}} = V + \frac{1}{2}\mathbf{\sigma}\cdot{\hat{\mathbf{p}}} \mathcal{K} \mathbf{\sigma}\cdot{\hat{\mathbf{p}}},
\end{equation}

where $\mathcal{K}$ effectively determines the local influence of relativity on the system (Fig.~\ref{fig:plotK}),

\begin{equation}
  \mathcal{K} = \frac{2c^2}{2c^2 - V}.
\end{equation}

It is known that ZORA describes valence states in many-electron systems well \cite{lenthe_phd,sadlej_four_1995}, while describing core states less well. However, as valence states are the main contributors to J-coupling it should provide an appropriate level of theory for the present work.

Substituting the canonical momentum for a magnetic vector potential $\mathbf{A}$, $\hat{\mathbf{p}} \rightarrow \pi = \hat{\mathbf{p}} + \mathbf{A}$, and expanding we obtain the ZORA Hamiltonian in a magnetic field:

\begin{subequations}
\begin{align}
  \hat{H}^{\textrm{ZORA}} &= \hat{V} + \frac{1}{2}(\hat{\mathbf{p}}\mathcal{K}\hat{\mathbf{p}} + i \sigma(\hat{\mathbf{p}}\mathcal{K})\times\hat{\mathbf{p}} \label{eqn:zoraEM1} \\
                          &+ \hat{\mathbf{p}}\mathcal{K}\mathbf{A} + \mathbf{A}\mathcal{K}\hat{\mathbf{p}} + i\sigma \left[\hat{\mathbf{p}}\times(\mathcal{K}\mathbf{A}) + \mathbf{A}\times(\mathcal{K}\hat{\mathbf{p}})\right] \label{eqn:zoraEM2}\\
                          &+ \mathcal{K}\mathbf{A}\cdot\mathbf{A}) \label{eqn:zoraEM3}
\end{align}
\end{subequations}

It can be observed that for $\mathcal{K}=1$ the ZORA Hamiltonian reduces to the nonrelativistic Levy-Leblond Hamiltonian \cite{levy-leblond_galilean_1967} plus spin-orbit coupling. The right hand side of \ref{eqn:zoraEM1} corresponds to the EM-free ZORA Hamiltonian. We concentrate on the scalar-relativistic terms, parts \ref{eqn:zoraEM2} and \ref{eqn:zoraEM3}, so we neglect the third term of \ref{eqn:zoraEM1}, representing spin-orbit coupling. However, we note that the effect of spin-orbit coupling on J-coupling can be significant in some compounds \cite{autschbach_nuclear2_2000,moncho_relativistic_2010}.

\subsection{\label{sec:theory-j}NMR J-coupling}

In NMR, indirect spin-spin coupling or J-coupling is an interaction between two nuclear moments due to indirect coupling mediated by the electrons in the system. The first analysis of this interaction came with Ramsey and Purcel\cite{ramsey_interactions_1952} and Ramsey\cite{ramsey_electron_1953}, who decomposed the interaction into four mechanisms: two due to interactions of the electron spins with the nuclear moments and two due to electron currents induced by the nuclear moments. When spin-orbit coupling is neglected these can be treated separately.

Following Ramsey's second-order perturbation analysis, the reduced spin coupling tensor, $K^{AB}$ between nuclei A and B, can be expressed as a second derivative of the system energy with respect to the two interacting nuclear moments, $\bmu = \gamma \hbar \mathbf{I}$, where $\gamma$ is the nucleus' gyromagnetic ratio and $\mathbf{I}$ is the nucleus' spin:

\begin{equation}
  K^{AB} = \left. \frac{\partial^2E}{\partial\bmu_A\partial\bmu_B} \right|_{\bmu_A=0,\bmu_B=0}
\end{equation}

We can then express the observed $J$ tensor in terms of the reduced spin coupling tensor, $J^{AB} = \frac{\hbar}{2\pi} \gamma_A \gamma_B K^{AB}$.

For our system of nuclear dipole moments the magnetic vector potential is, in the symmetric gauge with ($\mathbf{r}_{N} = \mathbf{r} - \mathbf{R}_N$), $\mathbf{A} = \sum_N \alpha^2 \frac{\bmu_{N} \times \mathbf{r}_{N}}{|\mathbf{r}_{N}|^3}$, where $\alpha$ is the fine-structure constant. Determining the derivatives of the ZORA Hamiltonian in this $\mathbf{A}$-field with respect to the interacting nuclear magnetic moments will allow us to use second-order perturbation theory to calculate $K^{AB}$.

We will use superscripts to represent order of perturbation with respect to the perturbation parameters $\bmu_A$ and $\bmu_B$:

\begin{equation}
  \hat{H}^{(n,m)} = \left. \left(\frac{\partial}{\partial \bmu_A}\right)^n\left(\frac{\partial}{\partial \bmu_B}\right)^m\hat{H}\right|_{\bmu_A=0,\bmu_B=0}
\end{equation}

Autschbach\cite{autschbach_nuclear_2000} obtained the following derivatives of the ZORA Hamiltonian, along with their equivalent $\hat{H}^{(0,1)}$ derivatives, as follows:

\begin{subequations}
\begin{align}
  \hat{H}^{(1,1)}_{\textrm{Z-dia};i,j} &= \mathcal{K}\alpha^4 \frac{\delta_{ij}(\mathbf{r}_A \cdot \mathbf{r}_B) - \mathbf{r}_{A;i} \mathbf{r}_{B;j}}{|\mathbf{r}_A|^3|\mathbf{r}_B|^3} \\
                                   &= \mathcal{K} H^{(1,1)}_{\textrm{dia}} \label{eqn:partZ-dia}
\end{align}
\end{subequations}

\begin{subequations}
\begin{align}
  \hat{H}^{(1,0)}_{\textrm{Z-para};i} &= \frac{\alpha^2}{2i} \left[\frac{\mathcal{K}}{|\mathbf{r}_A|^3} (\mathbf{r}_A \times \nabla)_i + (\mathbf{r}_A \times \nabla)_i \frac{\mathcal{K}}{|\mathbf{r}_A|^3}\right] \\
                                    &= \frac{\alpha^2}{i} \frac{\mathcal{K}}{|\mathbf{r}_A|^3} (\mathbf{r}_A \times \nabla)_i \\
                                    &= \mathcal{K} \hat{H}^{(1,0)}_{\textrm{para}}\label{eqn:partZ-para} 
\end{align}
\end{subequations}

\begin{subequations}
\label{eqn:partZ-spin}
\begin{align}
  \hat{H}^{(1,0)}_{\textrm{Z-spin};i} &= \frac{8\pi\alpha^2}{6} \mathcal{K} \sigma_i\delta(\mathbf{r}_A) \label{eqn:partZ-spinFC} \\
                                    &+ \frac{\mathcal{K}\alpha^2}{2}\left(\frac{3(\sigma\cdot\mathbf{r}_A)\mathbf{r}_{A;i}}{|\mathbf{r}_A|^5} - \frac{\sigma_i}{|\mathbf{r}_A|^3}\right) \label{eqn:partZ-spinSD} \\
                                    &+ \frac{\alpha^2}{2}\frac{1}{|\mathbf{r}_A|^3} \left[\{(\nabla \mathcal{K})\cdot \mathbf{r}_A\}\sigma_i - (\nabla_i \mathcal{K})(\sigma\cdot \mathbf{r}_A)\right] \label{eqn:partZ-spinZFC} 
\end{align}
\end{subequations}

We can choose to split the spin term into a spin-dipole analogue, $\mathcal{K}\hat{H}^{(1,0)}_{\textrm{SD}}$, plus a Fermi-contact analogue:

\begin{subequations}
\begin{align}
\hat{H}^{(1,0)}_{\textrm{Z-FC};i} &= \frac{8\pi\alpha^2}{6} \mathcal{K} \sigma_i\delta(\mathbf{r}_A) \\
                                &+ \frac{d\mathcal{K}}{dr}\frac{\alpha^2}{2}\left(\frac{(\sigma\cdot\mathbf{r}_A)\mathbf{r}_{A;i}}{|\mathbf{r}_A|^4}- \frac{\sigma_i}{|\mathbf{r}_A|^2}\right).
\end{align}
\end{subequations}

All the terms reduce to the non-relativistic versions \cite{joyce_first_2007} in the limit $c \to \infty$, that is, $\mathcal{K}=1$ and $\nabla\mathcal{K}=0$. We see changes for the relativistic case, $\mathcal{K} \ne 1$, especially in the case of the Eqn.~\ref{eqn:partZ-spin} where term \ref{eqn:partZ-spinZFC} is non-zero and the term \ref{eqn:partZ-spinFC} (equivalent to the Fermi-contact term) is zero for a nuclear Coulomb potential, neglecting finite-size nucleus effects. Finite-size nucleus effects have been found \cite{autschbach_magnitude_2009} to be somewhat significant in certain circumstances, contributing a 10 to 15\% reduction in isotropic J-couplings between heavy elements in the sixth row of the periodic table and light elements, and larger between two heavy elements.

We can now proceed to calculate $K$ by separating it into contributions from the interactions of the nuclear magnetic moments mediated by the electron spin, $K^{\textrm{Z-spin}}$, Eqn.~\ref{eqn:partZ-spin}, and from the interactions of the nuclear magnetic moments mediated by the electron charge current, $K^{\textrm{Z-orb}}$, Eqn.~\ref{eqn:partZ-dia} and \ref{eqn:partZ-para}.

\subsection{\label{sec:theory-spinmag}Spin magnetization density}

To calculate the contribution from the electron spin density we first find an expression for the magnetization density induced by a perturbing nuclear moment. From this we can compute the magnetic field induced at all of the other nuclei in the system.

We take the wave function to be a product of spin-restricted independent electron orbitals $\psi$, with a spin index $\sigma$. The magnetization density induced by a nuclear magnetic moment aligned along the $\mathbf{u}_j$ direction, $\mathbf{m}_j$, is calculated by summing the magnetization density induced when applying the perturbation with the spin quantized along each direction $\mathbf{u}_k$.

\begin{equation}
  \mathbf{m}_j(\mathbf{r}) = \sum_k m_{jk}(\mathbf{r}) \mathbf{u}_k
\end{equation}

where $m_{jk}(\mathbf{r})$ is given by

\begin{equation}
  m_{jk}(\mathbf{r}) = 4g\beta \sum_o \langle \psi^{(1)}_{o\uparrow jk} | \mathbf{r} \rangle \langle \mathbf{r} | \psi^{(0)}_{o\uparrow} \rangle,
\end{equation}

and the first order response in the wave function, $| \psi^{(1)}_{o\sigma jk} \rangle$, is given by

\begin{equation}
  \label{eqn:selfconspin}
  | \psi^{(1)}_{o\sigma jk} \rangle = \mathcal{G}(\epsilon) | \psi^{(0)}_{o\sigma} \rangle.
\end{equation}

$\mathcal{G}(\epsilon)$ is Green's function,

\begin{equation}
  \mathcal{G}(\epsilon) = \sum_e \frac{| \psi^{(0)}_{e} \rangle \langle \psi^{(0)}_{e} | }{\epsilon_o - \epsilon_e} (\hat{H}^{(0,1)}_{\textrm{Z-spin};j} + \hat{H}^{(1)}_{\textrm{xc}}),
\end{equation}

where $\epsilon_o$ and $\epsilon_e$ are the eigenvalues of the occupied and empty bands respectively, $\sum_e$ is a sum over empty bands and $\hat{H}^{(1)}_{\textrm{xc}}$ is the self-consistent first order variation in the Kohn-Sham exchange-correlation potential due to the first order change in the spin density:

\begin{equation}
\label{eqn:selfconstxcspin}
  \hat{H}^{(1)}_{\textrm{xc}} = V^{(1)}_{\rm xc} [m^{(1)}],
\end{equation}
 
where $m^{(1)}$ is the first order magnetization density.

As there is a first-order change in the Hamiltonian, we solve Eqn.~\ref{eqn:selfconspin} self-consistently by iteration.

Given the first order variation in the wave functions, the full expression for the spin term contribution to $K$ is, with an implicit rotation over the spin axes,

\begin{equation}
  K^{\textrm{Z-spin}}_{ij} =  2\textrm{Re} \sum_{o,\sigma} \int d\mathbf{r} \: \psi^{(0)}_{o}(\mathbf{r})^\dagger \hat{H}^{(1,0)}_{\textrm{Z-spin};i} \psi^{(1)}_{o\sigma j}(\mathbf{r}).
\end{equation}

We can rearrange this expression to get the effective spin coupling in terms of the induced magnetisation, including an implicit rotation over the spin axes,

\begin{widetext}
\begin{equation}
 K^{\textrm{Z-spin}}_{ij} = \alpha^2 \int d\mathbf{r} \: \left[ \mathcal{K} \left(\frac{3\mathbf{r}_{A;i} \mathbf{r}_{A;j} - |\mathbf{r}_A|^2 \delta_{ij}}{|\mathbf{r}_{A}|^5}\right) - \frac{d\mathcal{K}}{dr} \frac{\mathbf{r}_{A;i} \mathbf{r}_{A;j} - |\mathbf{r}_{A}|^2\delta_{ij}}{|\mathbf{r}_A|^4} \right] \mathbf{m}_j(\mathbf{r}).
\end{equation}
\end{widetext}

\subsection{\label{sec:theory-current}Current density}

To calculate the contribution from the electron charge current density to $K$ we first find an expression for the current density induced by a perturbing nuclear moment and subsequently the magnetic field induced at the receiving magnetic moment. As with the induced spin-magnetization density, we can calculate the induced magnetic field at all the receiving nuclei from this induced current density.

To first order the current perturbation term, $\hat{H}^{(0,1)}_{\textrm{Z-para},j}$, does not modify the magnetization density or the charge density so there is no first order change in the self-consistent potential. The first order variation in the orbitals is therefore

\begin{equation}
\label{eqn:selfconcurrent}
  | \psi^{(1)}_{o j} \rangle = \mathcal{G}(\epsilon) | \psi^{(0)}_{o} \rangle,
\end{equation}

where $\mathcal{G}(\epsilon)$ is Green's function,

\begin{equation}
  \mathcal{G}(\epsilon) = \sum_e \frac{| \psi^{(0)}_{e} \rangle \langle \psi^{(0)}_{e} | }{\epsilon_o - \epsilon_e} \hat{H}^{(0,1)}_{\textrm{Z-para};j},
\end{equation}

in which $\epsilon_o$ and $\epsilon_e$ are the eigenvalues of the occupied and empty bands respectively and $\sum_e$ is a sum over empty bands.

From second order perturbation theory we can write down $K^{\textrm{Z-orb}}$:

\begin{align}
\begin{split}
\label{eqn:orb2ndperturb}
  K^{\textrm{Z-orb}}_{ij} =  \sum_{o,\sigma} \int d\mathbf{r} \: & \psi^{(0)}_{o}(\mathbf{r})^\dagger \hat{H}^{(1,0)}_{\textrm{Z-para};i} \psi^{(1)}_{o j}(\mathbf{r}) + c.c. \\
                          + & \psi^{(0)}_{o}(\mathbf{r})^\dagger \hat{H}^{(1,1)}_{\textrm{Z-dia},ij} \psi^{(0)}(\mathbf{r}).
\end{split}
\end{align}

From the first order variation in the orbitals we can re-arrange Eqn.~\ref{eqn:orb2ndperturb} to find the induced current:

\begin{align}
  \mathbf{j}^{(1)}_j(\mathbf{r}) &= 2\sum_o [ 2 \Re \langle \psi^{(0)}_o | \mathbf{J}^{\textrm{Z-P}}(\mathbf{r})  | \psi^{(1)}_{oj} \rangle \\
   & + \langle \psi^{(0)}_o | \mathbf{J}_j^{\textrm{Z-D}}(\mathbf{r})  | \psi^{(0)}_o \rangle ],\nonumber
\end{align}

where our modified paramagnetic and diamagnetic current operators have the form

\begin{align}
  \begin{split}
  \mathbf{J}^{\textrm{Z-P}}(\mathbf{r})   &= -\mathcal{K} (\hat{\mathbf{p}}|\mathbf{r}\rangle\langle\mathbf{r}| + |\mathbf{r}\rangle\langle\mathbf{r}|\hat{\mathbf{p}}) / 2 \\
                                          &= \mathcal{K} \mathbf{J}^{\textrm{P}}(\mathbf{r})
  \end{split}\\
  \begin{split}
  \mathbf{J}_j^{\textrm{Z-D}}(\mathbf{r}) &= -\mathcal{K} \alpha^2 \frac{\mathbf{r}_{B} \times \hat{\mu}^j}{|\mathbf{r}_{kB}|^3} \\
                                          &= \mathcal{K} \mathbf{J}^{\textrm{D}}(\mathbf{r}),
  \end{split}
\end{align}

and so the orbital current contribution to $K$ is simply the Biot-Savart law acting on the induced current:

\begin{equation}
  K^{\textrm{Z-orb}}_{ij} = \alpha^2 \int d\mathbf{r} \: \left( \mathbf{j}^{(1)}_{j}(\mathbf{r}) \times \frac{\mathbf{r}_A}{|\mathbf{r}_A|^3}\right)_i.
\end{equation}

\subsection{\label{sec:theory-paw}Projector augmented wave}

The expressions derived in Sections \ref{sec:theory-spinmag} and \ref{sec:theory-current} cannot be directly applied in a pseudopotential based formalism. The use of pseudopotential implies that the valence wave functions have a non-physical form in the region close to the nucleus; the all-electron operators used in Sections  \ref{sec:theory-spinmag} and \ref{sec:theory-current}  are sensitive to the precise form of the wavefunctions near to the nuclei. The standard approach to deal with this problem is the projector augmented wave (PAW) formalism introduced by Van de Walle and Bl\"ochl \cite{van_de_walle_first-principles_1993,blochl_projector_1994}. PAW provides a practical way to transform an operator acting on the all-electron wavefunction (AE) $\wfae$ into an operator acting on the pseudo-wavefunction (PS) $\wfps$, allowing pseudopotentials to be used in calculations of properties that are sensitive to the form of the wavefunction near the nucleus. In particular, PAW proposes a linear transformation $\pawT$,

\begin{equation}
  \pawT = 1 + \sum_i(| \phi_i \rangle - | \tilde{\phi_i} \rangle)\langle p_i |,
\end{equation}

such that $\pawT \wfps = \wfae$ where $\phi_i$ and $\tilde{\phi_i}$ are atomic-like AE and PS partial waves at each atomic site for the spherically symmetric atom and pseudo-atom. The AE and PS partial waves form a complete basis within the augmentation region ($r<r_c$) and are equal outside. The functions $p_i$ are the corresponding projectors to $\tilde{\phi_i}$, defined such that $\langle p_i | \tilde{\phi}_j \rangle = \delta_{ij}$ and that they vanish outside the augmentation region.

So, for an AE operator $\hat{O}$,

\begin{equation}
  \langle \psi | \hat{O} | \psi \rangle = \langle \tilde{\psi} | \pawT^{\dagger} \hat{O} \pawT | \tilde{\psi} \rangle,
\end{equation}

we can derive the equivalent pseudo-operator $\widetilde{O} = \pawT^{\dagger} \hat{O} \pawT$. The explicit form of $\widetilde{O}$ is

\begin{equation}
  \widetilde{O} = \hat{O} + \sum_{ij} | p_i \rangle (\langle \phi_i | \hat{O} | \phi_j \rangle - \langle \tilde{\phi_i} | \hat{O} | \tilde{\phi_j} \rangle) \langle p_j |.
  \label{eqn:PAWoptrans}
\end{equation}

This allows us to calculate all-electron properties from pseudopotential calculations. Bl\"ochl notes that there is a degree of freedom in Eqn.~\ref{eqn:PAWoptrans} to add a term  of the form

\begin{equation}
\hat{B} - \sum_{ij} | p_i \rangle \langle \tilde{\phi_i} | \hat{B} | \tilde{\phi_j} \rangle \langle p_j |,
\label{eqn:Bfree}
\end{equation}

where $\hat{B}$ is an arbitrary local operator acting solely within the augmentation region. We might describe an all electron operator $\hat{O}$ in terms of a `fictitious' operator $\hat{O}_{\textrm{fict}}$ and a `real' operator $\hat{O}_{\textrm{real}}$, which are equal outside the augmentation region:

\begin{equation}
  \hat{O} = \hat{O}_{\textrm{fict}} + f(r) ( \hat{O}_{\textrm{real}} - \hat{O}_{\textrm{fict}}),
\end{equation}

where $f(r)$ is a cutoff function which is unity for $r<r_c$ and zero otherwise and so $\hat{B} =\hat{O}_{\textrm{real}} - \hat{O}_{\textrm{fict}}$. If we apply the PAW operator transform to $\hat{O}$ (Eqn.~\ref{eqn:PAWoptrans}) and add Eqn.~\ref{eqn:Bfree} we find:

\begin{equation}
  \widetilde{O} = \hat{O}_{\textrm{fict}} + \sum_{ij} | p_i \rangle (\langle \phi_i | \hat{O}_{\textrm{real}} | \phi_j \rangle - \langle \tilde{\phi_i} | \hat{O}_{\textrm{fict}} | \tilde{\phi_j} \rangle) \langle p_j |.
  \label{eqn:PAWfict}
\end{equation}

This allows us to substitute a new operator $\hat{O}_{\textrm{fict}}$, which is easier to numerically represent, for the pseudo calculation, so long as it is equal to the real operator, $\hat{O}_{\textrm{real}}$, outside the augmentation region, and perform the correction in the PAW augmentation. Bl\"ochl goes on to use this freedom to set pseudopotential theory within the PAW formalism by substituting the real nuclear Coulomb potential of an atom for a smoothed fictitious potential that is more amenable for evaluation in a plane wave basis set. 

We can use this technique to better represent the radially divergent part of the magnetic dipole perturbation, $\frac{1}{r^3}$, which when na\"ively applied in real space is ill-represented on the cartesian grid, leading to varying numerical predictions depending on the relative position of the perturbing nucleus to the real space grid. We replace it with a smoothed term,

\begin{equation}
  \hat{O}_{\textrm{fict}} = \frac{1 - \exp(-(r/r_0)^3)}{r^3},
\end{equation}

which is well represented on a coarser real-space grid and make the correction using Eqn.~\ref{eqn:PAWfict} during PAW augmentation.

\begin{figure}
 \includegraphics[width=0.75\linewidth]{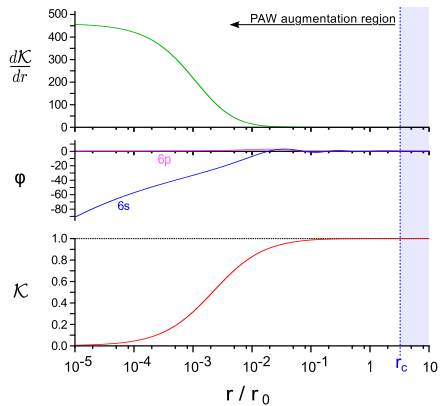}
 \caption{\label{fig:plotK}Plot of $\mathcal{K}(r)$, the 6s and 6p all-electron partial waves $\phi$, and $\frac{d\mathcal{K}}{dr}$ for a scalar-relativistic lead atom. Note that $\mathcal{K}(r)$ quickly reaches unity and $\frac{d\mathcal{K}}{dr}$ quickly reaches zero for $r \ll r_c = 2.36 a_0$, the pseudopotential cut off radius. Also note that only the 6s orbital has significant character in the region where $\frac{d\mathcal{K}}{dr} \gg 0$ while the 6p has none, meaning the behaviour of the Z-FC term will be similar to the non-relativistic FC term.}
\end{figure}

\subsection{\label{sec:theory-jpaw}J-coupling with PAW}

Unlike the calculation of NMR magnetic shielding, which requires the use of GIPAW to preserve translational invariance in a magnetic field \cite{pickard_all-electron_2001}, the calculation of J-coupling can use a gauge fixed on the perturbing atom and so only standard PAW is required.

We will now apply PAW to the derived ZORA-level scalar-relativistic theory of J-coupling to get pseudo-operators in terms of modifications to the non-relativistic operators. The diamagnetic and paramagnetic current operators are simple multiples of $\mathcal{K}$, while the Fermi-contact term disappears for a nuclear Coulomb potential and the spin-dipole term turns into a simple multiple of $\mathcal{K}$ plus a Fermi-contact-like term which is sharply localised by the gradient of $\mathcal{K}$ (Figure \ref{fig:plotK}). As the region where $\mathcal{K}\neq 1$ is localised inside the augmentation region, the operators are equal to their non-relativistic equivalents outside, and so we can apply all these corrections accurately in PAW augmentation for no extra computational cost \cite{yates_relativistic_2003} using Eqn.~\ref{eqn:PAWfict}.

In the case of the spin contribution, the sharply localised FC/Z-FC perturbation operator can be written as 

\begin{align}
  \widetilde{H}^{(1,0)}_{\textrm{Z-FC}} &= \sum_{\mathbf{R},n,m} | p_{\mathbf{R},n} \rangle \langle \phi_{\mathbf{R},n} |\hat{H}^{(1,0)}_{\textrm{Z-FC}} | \phi_{\mathbf{R},m} \rangle \langle p_{\mathbf{R},m} |,
\end{align}

where we have used the on-site approximation. The SD/Z-SD PAW augmentation is also evaluated on a real radial grid, while the un-augmented contribution is calculated in Fourier space on the augmented magnetization density.

\begin{equation}
  \widetilde{H}^{(1,0)}_{\textrm{Z-SD}} = \hat{H}^{(1,0)}_{\textrm{SD}} + \Delta \hat{H}^{(1,0)}_{\textrm{Z-SD}},
\end{equation}

where, using the on-site approximation again,

\begin{align*}
  \Delta \hat{H}^{(1,0)}_{\textrm{Z-SD}} = \sum_{\mathbf{R},n,m} | p_{\mathbf{R},n} \rangle (&\langle \phi_{\mathbf{R},n} | \mathcal{K}\hat{H}^{(1,0)}_{\textrm{SD}} | \phi_{\mathbf{R},m} \rangle \\
  - &\langle \tilde{\phi}_{\mathbf{R},n}| \hat{H}^{(1,0)}_{\textrm{SD}} | \tilde{\phi}_{\mathbf{R},m} \rangle) \langle p_{\mathbf{R},m} |. 
\end{align*}

The total spin contribution to the reduced coupling tensor is then

\begin{align}
  K^{\textrm{Z-spin}} &= \tilde{K}^{\textrm{Z-spin}} + \Delta K^{\textrm{Z-spin}} \\
  \tilde{K}^{\textrm{Z-spin}} &= -\frac{\mu_0}{3} \int d \mathbf{G} \: \left[ \frac{3(\mathbf{\tilde{m}}^{(1)}(\mathbf{G}) \cdot \mathbf{G})\mathbf{G} - \mathbf{\tilde{m}}^{(1)}(\mathbf{G}) G^2}{G^2} \right] e^{i\mathbf{G}\cdot\mathbf{r}_A} \label{eqn:sdbare} \\
  \Delta K^{\textrm{Z-spin}} &= \sum_{oo^\prime\sigma\sigma^\prime} \langle \psi^{(0)}_{o\sigma} | \widetilde{H}^{(1,0)}_{\textrm{Z-FC}} + \Delta \hat{H}^{(1,0)}_{\textrm{SD}} | \psi^{(1)}_{o^\prime\sigma^\prime} \rangle,
\end{align}

where $\mathbf{\tilde{m}}^{(1)}(\mathbf{G})$ is the un-augmented spin density in reciprocal space, to which we apply the spin dipole operator and slow Fourier transform at the position of the receiving nucleus



The diamagnetic current operators do not receive PAW augmentation, as we cannot make the off-site approximation due to the complexity of calculating the relevant matrix elements, and so only contributes to the bare current. This has the effect of ignoring the diamagnetic augmented current for off-site nuclei, though the diamagnetic contribution is the smallest J-coupling component and may only be really relevant for the calculation of anisotropic couplings \cite{sychrovsky_nuclear_2000}.



The PAW augmented paramagnetic operator is

\begin{align}
  \tilde{H}^{(1,0)}_{\textrm{Z-para}} &= \hat{H}^{(1,0)}_{\textrm{Z-para}} + \Delta \hat{H}^{(1,0)}_{\textrm{Z-para}} \\
  \Delta \hat{H}^{(1,0)}_{\textrm{Z-para}} &= \sum_{\mathbf{R},n,m} | p_{\mathbf{R},n} \rangle (\langle \phi_{\mathbf{R},n} | \mathcal{K} \hat{H}^{(1,0)}_{\textrm{para}} | \phi_{\mathbf{R},m} \rangle \\
  &- \langle \tilde{\phi}_{\mathbf{R},n} | \hat{H}^{(1,0)}_{\textrm{para}} | \tilde{\phi}_{\mathbf{R},m} \rangle) \langle p_{\mathbf{R},m} |.
\end{align}

The bare contribution of the paramagnetic current is calculated in Fourier space on the pseudo-current density and is PAW augmented on a real radial grid with an on-site approximation, justified by the short-rangeness of the interaction. The total orbital current contribution is then

\begin{align}
  K^{\textrm{Z-orb}} &= \tilde{K}^{\textrm{Z-orb}} + \Delta K^{\textrm{Z-orb}} \\
  \tilde{K}^{\textrm{Z-orb}} &= \mu_0 \int d \mathbf{G} \: \frac{i \mathbf{G} \times \mathbf{\tilde{j}}^{(1)}(\mathbf{G})}{G^2} e^{i\mathbf{G}\cdot\mathbf{r}_A} \\
  \Delta K^{\textrm{Z-orb}} &= 2 \Re  \sum_{oo^\prime\sigma} \langle \psi^{(0)}_{o\sigma} | \Delta \hat{H}^{(1,0)}_{\textrm{Z-para}} | \psi^{(1)}_{o^\prime\sigma} \rangle,
\end{align}

where $\mathbf{\tilde{j}}^{(1)}(\mathbf{G})$ is the un-augmented induced current density, paramagnetic and diamagnetic, in Fourier space, to which we apply the Biot-Savart law in reciprocal space and slow Fourier transform at the position of the receiving nucleus.

The total indirect coupling tensor between the two nuclei is then the sum of $K^{\textrm{Z-spin}}$ and $K^{\textrm{Z-orb}}$.

\subsection{\label{section:extrauspconsid}Additional considerations for ultrasoft pseudopotentials}

The so-called `ultrasoft' pseudopotential formalism introduced by Vanderbilt\cite{vanderbilt_soft_1990} is the most computationally efficient form of pseudopotential generally providing numerically converged results with significantly smaller basis-sets. This is particularly important for elements which require semi-core states to be treated as valence for accurate results e.g. `3p' states in the 3d transition metal series. While ultrasoft potentials are efficient from a user's point of view, there is some additional complexity when implemented in an electronic-structure code.  The key ingredient of the ultrasoft scheme is that the norm of the the pseudo partial-waves in the augmentation region is different from that of the corresponding all-electron partial-waves. We can thus define a non-zero charge augmentation term ${\rm Q}_{{\bf R},nm}({\mathbf{r}})$:


\begin{equation}\label{eqn:Qfr}
  {\rm Q}_{{\bf R},nm}({\mathbf{r}}) = \langle \phi_{{\bf R},n} |{\bf r}\rangle\langle{\bf r}| \phi_{{\bf R},m} \rangle - \langle \tilde{\phi}_{{\bf R},n} |{\bf r}\rangle\langle{\bf r}| \tilde{\phi}_{{\bf R},m} \rangle. 
\end{equation}

The norm of a pseudo-wave function can be computed as the expectation value of the pseudo operator $\tilde{1}=S$.  Using Eqn.~\ref{eqn:Qfr},

\begin{equation}\label{eqn:overlap}
  S = 1 + \sum_{{\bf R},n,m} |p_{{\bf R},n} \rangle {\rm q}_{{\bf R},nm} \langle p_{{\bf R},m} |
\end{equation} 

where

\begin{equation}\label{eqn:qfr}
  {\rm q}_{{\bf R},nm} = \langle \phi_{{\bf R},n} | \phi_{{\bf R},m} \rangle-\langle \tilde{\phi}_{{\bf R},n} | \tilde{\phi}_{{\bf R},m} \rangle. 
\end{equation} 

As a result, a normalized eigenstate of the pseudo Hamiltonian obeys the generalized equations:

\begin{equation}\label{eqn:ham-zero}
  \tilde{H}\ket{\tilde{\psi}_{o} } = \varepsilon_{o}S\ket{\tilde{\psi}_{o}},
\end{equation}

and

\begin{equation}\label{eqn:genorth}
  \bra{\tilde{\psi}_{o}}S\ket{\tilde{\psi}_{o^\prime}}=\delta_{o,o^\prime}.
\end{equation}

The pseudo-Hamiltonian $\tilde{H}$ can be derived using Eqn.~\ref{eqn:PAWoptrans} as

\begin{equation}
  \tilde{H} = -\nabla^2 + V_{\textrm{eff}} + \sum_{{\bf R},n,m} |p_{{\bf R},n} \rangle {\rm D}_{{\bf R},nm} \langle p_{{\bf R},m} |
\end{equation} 

where $ {\rm D}_{{\bf R},nm}$ is given by

\begin{equation}
  {\rm D}_{{\bf R},nm}= {\rm D}_{{\bf R},nm}^{0} + \int d{\mathbf{r} \:} V_{\textrm{eff}}({\bf r}){\rm Q}_{{\bf R},nm}({\bf r}).
\end{equation}

${\rm D}_{{\bf R},nm}^{0}$ is obtained from the construction of the pseudopotential\cite{laasonen_car-parrinello_1993}, and $V_{\textrm{eff}}$ is the screened local potential. We note that the norm-conserving pseudopotential scheme can be regarded as special case in which ${\rm q}_{{\bf R},nm}=0$ by definition, and by convention the terms ${\rm Q}_{{\bf R},nm}$ are assumed to vanish. The charge density in the ultrasoft scheme is given by

\begin{align}
  n(\mathbf{r}) = &\sum_{o\sigma} [ \psi^{*(0)}_{o\sigma}(\mathbf{r}) \psi^{(0)}_{o\sigma}(\mathbf{r}) \\
                  &+ \sum_{\mathbf{R},n,m} Q_{\mathbf{R},nm}(\mathbf{r}) \langle p_{\mathbf{R},n} | \psi^{(0)}_{o\sigma} \rangle \langle \psi^{(0)}_{o\sigma} | p_{\mathbf{R},m} \rangle ]
\label{eqn:uspcharge}
\end{align}

In practice $Q_{\mathbf{R},nm}$ and hence $n(\mathbf{r})$ would be prohibitively expensive to represent in a planewave basis, and so $Q_{\mathbf{R},nm}$ is replaced by a pseudized augmentation charge $\tilde{Q}_{\mathbf{R},nm}$, where the pseudization conserves the electrostatic moments of the charge\cite{laasonen_car-parrinello_1993}. The grid-representable charge density  $\bar{n}(\mathbf{r})$ is given by

\begin{align}
  \bar{n}(\mathbf{r}) = &\sum_{o\sigma} [ \psi^{*(0)}_{o\sigma}(\mathbf{r}) \psi^{(0)}_{o\sigma}(\mathbf{r}) \\
                      &+ \sum_{\mathbf{R},n,m} \tilde{Q}_{\mathbf{R},nm}(\mathbf{r}) \langle p_{\mathbf{R},n} | \psi^{(0)}_{o\sigma} \rangle \langle \psi^{(0)}_{o\sigma} | p_{\mathbf{R},m} \rangle ]
\label{eqn:uspgridcharge}
\end{align}

The ZORA J-coupling operators derived in Section~\ref{sec:theory-jpaw} are still valid for ultrasoft potentials, as the PAW formulation made no assumption regarding the norm of the partial waves used in the PAW transformation. However, changes are required when computing the first order induced current and magnetization density. Firstly, we note that to avoid a costly sum over unoccupied states the first-order change in the wave functions is computed using a conjugate-gradient minimization\cite{joyce_first_2007}. For ultrasoft pseudopotentials the minimization routines must be adapted to allow for the generalised orthonormality condition, Eqn.~\ref{eqn:genorth}. This can be done straight-forwardly following the method outlined in Appendix B of Ref. \onlinecite{yates_calculation_2007}. Secondly, care must be taken in computing the induced magnetization density. Following the ultrasoft charge density in Eqn.~\ref{eqn:uspgridcharge} we introduce the magnetization density $\bar{m}^{(1)}$ 

\begin{align}
  \bar{m}^{(1)}(\mathbf{r}) =& 2\sum_o [ \psi^{*(1)}_o(\mathbf{r}) \psi^{(0)}_o(\mathbf{r}) + c.c.\\
&+ \sum_{\mathbf{R},n,m} \tilde{Q}_{\mathbf{R},nm}(\mathbf{r}) \left(\langle p_{\mathbf{R},n} | \psi^{(1)}_o \rangle \langle \psi^{(0)}_o | p_{\mathbf{R},m} \rangle + c.c.\right) ]
\end{align}

When using ultrasoft potentials we compute Eqn.~\ref{eqn:sdbare} using $\bar{m}^{(1)}(\mathbf{r})$ in order to capture off-site contributions to the spin-dipolar contribution. However, this requires that care is taken to subtract from every receiving atom the on-site contribution from the augmentation charge so as not to double count with the PAW on-site augmentation.

In addition, the self-consistent exchange-correlation term, Eqn.~\ref{eqn:selfconstxcspin}, should also be augmented to reflect the first order change in the $D$ matrix \cite{dal_corso_density-functional_2001}:


\begin{align}
\label{eqn:selfconstxcspinusp}
  \hat{H}^{(1)}_{\textrm{xc}} &= V^{(1)}_{\textrm{xc}} [\bar{m}^{(1)}] \\
    & + \sum_{\mathbf{R},n,m} |p_{{\bf R},n} \rangle \left[ \int d\mathbf{r} \: V^{(1)}_{\textrm{xc}}  [\bar{m}^{(1)}] (\mathbf{r}) \tilde{Q}_{\mathbf{R},nm}(\mathbf{r}) \right] \langle p_{{\bf R},m} |. \nonumber
\end{align}

\section{\label{sec:calculations}Calculations}

The PAW scalar-relativistic theory as proposed has been implemented in the CASTEP planewave DFT code\cite{CASTEP} using norm-conserving pseudopotentials \cite{troullier_efficient_1991} and ultrasoft pseudopotentials \cite{vanderbilt_soft_1990,laasonen_car-parrinello_1993}. In the case of norm-conserving pseudopotentials this involved modifying only the preparatory calculation of PAW matrix elements, as the pseudo-Hamiltonian remains the same as for the existing non-relativistic theory except for PAW augmentation. In the case of ultrasoft pseudopotentials modifications were needed to address the issues mentioned in Section \ref{section:extrauspconsid}. In addition, PAW is also used to smooth the dipole operator and reconstruct it in real space to provide greater numerical stability with respect to real grid spacing, as described in Eqn.~\ref{eqn:PAWfict}.

We will first seek to validate our implementation against all-electron, local orbital basis set quantum chemical predictions and experimental measurements of J-coupling for both molecules containing light atoms, to be treated non-relativistically, and molecules containing heavy atoms, to be treated relativistically. We will then proceed to use the implemented code to predict J-coupling in two large novel molecular crystals containing silver atoms.

\subsection{\label{sec:lantto}Validation against existing quantum chemistry: Non-relativistic}

\begin{figure}
 \centering
 \includegraphics[height=0.75\linewidth]{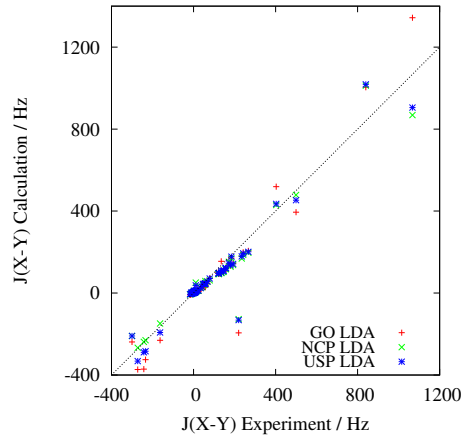}
 \caption{\label{fig:lda_v_exp}Comparison of isotropic J-couplings in a number of small molecules calculated using Gaussian orbital basis sets (red crosses), norm-conserving pseudopotentials (green diagonal crosses) and ultrasoft pseudopotentials (blue stars), all at the non-relativistic level of theory and using the LDA exchange-correlation functional, against experimental values. Gaussian orbital and experimental values are from Ref.~\onlinecite{lantto_spinspin_2002}.}
\end{figure}

\begin{figure}
 \centering
 \includegraphics[height=0.75\linewidth]{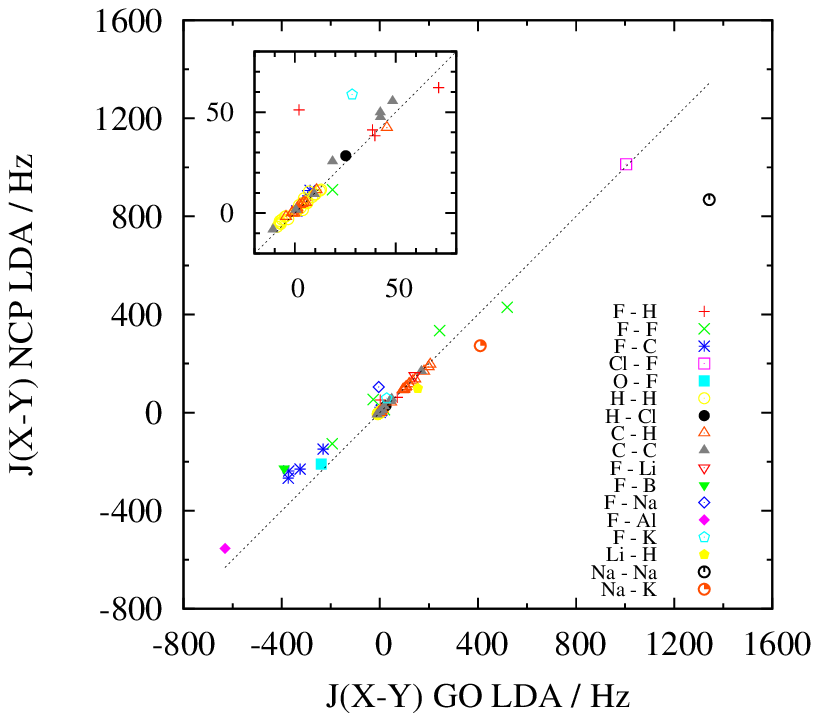}
 \caption{\label{fig:ncp_v_go}Comparison of isotropic J-couplings in a number of small molecules calculated using norm-conserving pseudopotentials against calculations using Gaussian orbital basis sets, both at the non-relativistic level of theory and using the LDA exchange-correlation functional. Gaussian orbital values are from Ref.~\onlinecite{lantto_spinspin_2002}.}
\end{figure}

\begin{figure}
 \centering
 \includegraphics[height=0.75\linewidth]{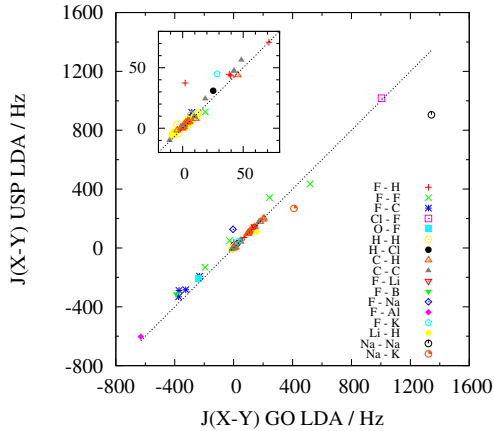}
 \caption{\label{fig:usp_v_go}Comparison of isotropic J-couplings in a number of small molecules calculated using ultrasoft pseudopotentials against calculations using Gaussian orbital basis sets, both at the non-relativistic level of theory and using the LDA exchange-correlation functional. Gaussian orbital values are from Ref.~\onlinecite{lantto_spinspin_2002}.}
\end{figure}

Lantto, Vaara and Helgaker \cite{lantto_spinspin_2002} provide a benchmark on 34 small molecules containing light atoms with a total of 60 couplings calculated using localised Gaussian orbital basis sets at the LDA, BLYP, B3LYP and MCSCF levels of theory along with some experimental values. Included in the light atoms is fluorine, which is notoriously difficult to treat well with density functional theory\cite{sadoc19f_2011}. We have run the same calculations at a non-relativistic level of theory using both norm-conserving and ultrasoft pseudopotentials with the LDA functional and present the results in Tables \ref{tab:nrel_benchmark_F}, \ref{tab:nrel_benchmark_N}, \ref{tab:nrel_benchmark_Na} and \ref{tab:nrel_benchmark_rest} and Figures \ref{fig:lda_v_exp}, \ref{fig:ncp_v_go} and \ref{fig:usp_v_go}.

As expected, the fluorine couplings perform poorly compared to experiment, particularly \js(F-C). In addition, they compare poorly to the Gaussian orbital basis set calculations. The majority of the difference between the pseudopotential calculations and the Gaussian orbital basis set calculations is from the Fermi-contact term contribution, suggesting that the origin of the disagreement might lie in the difficulty of constructing Gaussian basis sets with sufficient flexibility in the core to represent Fermi-contact coupling\cite{helgaker_basis-set_1998}, a problem that is circumvented with pseudopotentials when using the PAW transformed operator.

The root mean squared deviation (RMSD) agreements of, where available, experimental couplings with LDA Gaussian orbital, norm-conserving and ultrasoft calculations was 31.3, 23.5 and 22.7 Hz respectively. Excluding systems including fluorine gives RMSD agreements with experiment of 25.6, 20.9 and 17.6 Hz. Excluding systems involving sodium, the other significantly poorly performing element, gives RMSD agreements with experiment of 25.0, 19.2 and 20.0 Hz respectively. Excluding systems including both fluorine and sodium gives RMSD agreements of 11.5, 13.0 and 11.6 Hz respectively.

The RMSD agreements of Gaussian orbital LDA with norm-conserving and ultrasoft calculations was 42.2 and 36.4 Hz respectively, 42.4 and 39.4 Hz excluding fluorine systems, 23.5 and 14.9 Hz excluding sodium systems and 5.4 and 4.4 Hz excluding both fluorine and sodium systems.

This indicates that, by and large, norm-conserving and ultrasoft pseudopotential calculations give couplings favourably comparable to equivalent localised Gaussian orbital calculations and to experiment.

\begin{table*}[p!]
 \scriptsize
 \caption{\label{tab:nrel_benchmark_F}Non-relativistic benchmark on fluorine containing molecules. Gaussian orbital LDA, BLYP, MCSCF calculations and experimental numbers from Ref.~\onlinecite{lantto_spinspin_2002}. NCP LDA and USP LDA values calculated with the implementation in CASTEP. All values are isotropic J-coupling in Hz.}
 \begin{ruledtabular}
 \begin{tabular}{lllrrrrrr}
  Structure & Site A & Site B & GO LDA & GO BLYP & GO MCSCF & NCP LDA & USP LDA & Exp \\
  \hline
  BF & F1 & B1 & -390.4 & -382.8 & -222.4 & -227.5 & -312.7 &  \\
  \hline
  FHF$^{-}$ & F1 & F2 & -194.5 & -124.6 & -238.6 & -127.1 & -131.9 & 220.0 \\
   & F1 & H1 & 2.0 & 0.5 & 10.1 & 51.1 & 37.3 & 11.0 \\
  \hline
  ClF$_3$ & F1 & F1 & 519.1 & 558.7 & 404.0 & 429.4 & 435.0 & 403.0 \\
  \hline
  HF & H1 & F1 & 394.6 & 389.0 & 539.3 & 478.7 & 453.2 & 499.8 \\
  \hline
  C$_6$H$_4$F$_2$ & F1 & C1 & -375.6 & -392.7 & -210.7 & -245.6 & -295.3 & -242.6 \\
   & F1 & C2 & 21.3 & 28.8 & 33.1 & 25.9 & 27.8 & 24.3 \\
   & F1 & C3 & 6.8 & 8.1 & 7.5 & 8.7 & 9.5 & 8.2 \\
   & F1 & C4 & -0.7 & 0.3 & 4.6 & 0.1 & 0.3 & 2.7 \\
   & F1 & F2 & 6.4 & 7.5 & 11.2 & 6.6 & 8.3 & 17.4 \\
   & F1 & H1 & 7.9 & 7.9 & 1.1 & 8.6 & 9.4 & 7.9 \\
   & F1 & H2 & 2.3 & 3.4 & 6.8 & 2.7 & 2.9 & 4.5 \\
  \hline
  ClF & Cl1 & F1 & 1004.3 & 979.1 & 832.3 & 1012.4 & 1018.1 & 840.0 \\
  \hline
  OF$_2$ & O1 & F1 & -238.6 & -249.3 & -309.0 & -209.5 & -209.3 & -300.0 \\
  \hline
  AlF & F1 & Al1 & -630.7 & -636.6 & -627.1 & -553.8 & -603.1 &  \\
  \hline
  CHF$_3$ & F1 & C1 & -344.0 & -361.6 & -225.6 & -239.1 & -301.7 & -272.2 \\
   & F1 & F1 & -43.3 & -16.7 & 125.9 & 32.4 & 30.9 &  \\
   & F1 & H1 & 71.2 & 88.2 & 79.1 & 62.1 & 71.0 & 79.1 \\
   & C1 & H1 & 180.5 & 228.0 & 236.8 & 168.3 & 178.7 & 235.3 \\
  \hline
  CH$_2$F$_2$ & F1 & C1 & -325.6 & -340.2 & -229.0 & -233.4 & -284.1 & -233.9 \\
   & F1 & F2 & -36.8 & -3.3 & 140.0 & 45.9 & 54.4 &  \\
   & F1 & H1 & 39.6 & 54.3 & 52.0 & 38.3 & 43.4 & 48.6 \\
   & C1 & H1 & 134.3 & 173.8 & 175.7 & 128.4 & 135.3 & 180.4 \\
  \hline
  CH$_3$F & F1 & C1 & -293.9 & -306.3 & -212.4 & -216.0 & -255.9 & -163.0 \\
   & F1 & H1 & 38.4 & 49.7 & 48.6 & 41.2 & 44.6 & 46.4 \\
   & C1 & H1 & 108.3 & 141.8 & 141.5 & 105.4 & 110.5 & 147.3 \\
   & H1 & H1 & -4.4 & -7.1 & -11.5 & -1.5 & -1.4 &  \\
  \hline
 \end{tabular}
 \end{ruledtabular}
\end{table*}

\begin{table*}[p!]
 \scriptsize
 \caption{\label{tab:nrel_benchmark_N}Non-relativistic benchmark on nitrogen containing molecules. Gaussian orbital LDA, BLYP, MCSCF calculations and experimental numbers from Ref.~\onlinecite{lantto_spinspin_2002}. NCP LDA and USP LDA values calculated with the implementation in CASTEP. All values are isotropic J-coupling in Hz.}
 \begin{ruledtabular}
 \begin{tabular}{lllrrrrrr}
  Structure & Site A & Site B & GO LDA & GO BLYP & GO MCSCF & NCP LDA & USP LDA & Exp \\
  \hline
  HNC & C1 & H1 & 10.7 & 10.6 & 16.4 & 11.5 & 7.9 &  \\
  \hline
  HCONH$_2$ & C1 & H1 & 144.2 & 183.2 & 183.2 & 136.4 & 142.3 & 193.1 \\
   & C1 & H2 & 4.4 & 5.2 & 2.8 & 5.3 & 6.0 & 2.7 \\
   & C1 & H3 & -1.4 & -1.8 & -4.2 & 0.1 & 0.5 & -3.7 \\
   & H2 & H3 & 6.9 & 9.0 & 3.8 & 10.7 & 10.9 & 2.2 \\
   & H2 & H1 & 0.9 & 0.9 & 0.8 & 1.2 & 1.1 & 2.3 \\
   & H3 & H1 & 10.0 & 14.1 & 11.8 & 9.2 & 9.8 & 13.9 \\
  \hline
  HCN & C1 & H1 & 205.8 & 260.8 & 249.3 & 196.7 & 201.2 & 267.3 \\
  \hline
  NH$_3$ & H1 & H1 & -5.0 & -9.0 & -11.3 & -1.8 & 3.2 & -10.0 \\
  \hline
  CH$_3$NC & C1 & C2 & -10.0 & -10.8 & -5.2 & -6.9 & -8.4 &  \\
   & C1 & H1 & 105.4 & 139.1 & 143.5 & 102.7 & 107.0 & 145.2 \\
   & C2 & H1 & 1.1 & 2.6 & 2.6 & 1.5 & 1.3 & 2.7 \\
   & H1 & H1 & -7.5 & -11.6 & -19.1 & -3.9 & -4.0 &  \\
  \hline
  CH$_3$CN & C1 & C2 & 43.5 & 56.5 & 72.0 & 51.0 & 48.7 & 58.0 \\
   & C1 & H1 & 99.4 & 130.5 & 142.4 & 97.1 & 101.2 & 135.7 \\
   & C2 & H1 & -4.4 & -6.9 & -15.5 & -1.8 & -1.6 & -9.9 \\
   & H1 & H1 & -8.7 & -13.8 & -22.9 & -6.7 & -5.4 & -16.9 \\
  \hline
 \end{tabular}
 \end{ruledtabular}
\end{table*}

\begin{table*}[p!]
 \scriptsize
 \caption{\label{tab:nrel_benchmark_Na}Non-relativistic benchmark on lithium, sodium and potassium containing molecules. Gaussian orbital LDA, BLYP, MCSCF calculations and experimental numbers from Ref.~\onlinecite{lantto_spinspin_2002}. NCP LDA and USP LDA values calculated with the implementation in CASTEP. All values are isotropic J-coupling in Hz.}
 \begin{ruledtabular}
 \begin{tabular}{lllrrrrrr}
  Structure & Site A & Site B & GO LDA & GO BLYP & GO MCSCF & NCP LDA & USP LDA & Exp \\
  \hline
  LiF & F1 & Li1 & 141.1 & 160.0 & 192.9 & 152.3 & 145.5 & 172.3 \\
  \hline
  NaF & F1 & Na1 & -4.8 & 39.5 & 193.9 & 104.2 & 126.2 &  \\
  \hline
  LiH & Li1 & H1 & 155.0 & 223.6 & 152.7 & 98.7 & 109.0 & 134.9 \\
  \hline
  Na$_2$ & Na1 & Na2 & 1343.3 & 1375.3 & 1243.9 & 868.4 & 905.1 & 1067.2 \\
  \hline
  KF & F1 & K1 & 28.4 & 43.2 & 76.6 & 58.7 & 44.9 & 57.8 \\
  \hline
  KNa & Na1 & K1 & 409.5 & 401.3 & 480.0 & 272.7 & 269.0 &  \\
  \hline
 \end{tabular}
 \end{ruledtabular}
\end{table*}

\begin{table*}[p!]
 \scriptsize
 \caption{\label{tab:nrel_benchmark_rest}Non-relativistic benchmark on remaining molecules. Gaussian orbital LDA, BLYP, MCSCF calculations and experimental numbers from Ref.~\onlinecite{lantto_spinspin_2002}. NCP LDA and USP LDA values calculated with the implementation in CASTEP. All values are isotropic J-coupling in Hz.}
 \begin{ruledtabular}
 \begin{tabular}{lllrrrrrr}
  Structure & Site A & Site B & GO LDA & GO BLYP & GO MCSCF & NCP LDA & USP LDA & Exp \\
  \hline
  PH$_3$ & H1 & H1 & -7.7 & -12.5 & -12.9 & -5.0 & -4.8 & -13.4 \\
  \hline
  SiH$_4$ & H1 & H1 & 2.0 & 4.1 & 0.3 & 4.0 & 4.2 & 2.6 \\
  \hline
  H$_2$O & H1 & H2 & -3.3 & -6.4 & -9.4 & -3.0 & -4.1 & -7.3 \\
  \hline
  CH$_4$ & C1 & H1 & 99.8 & 132.7 & 120.8 & 93.5 & 96.4 & 120.9 \\
  \hline
  CH$_3$SiH$_3$ & C1 & H1 & 94.4 & 121.3 & 115.7 & 93.1 & 96.6 & 122.5 \\
   & C1 & H4 & 3.7 & 5.5 & 3.4 & 4.6 & 4.8 & 4.6 \\
   & H1 & H1 & -7.2 & -11.6 & 15.2 & -4.9 & -3.9 &  \\
   & H4 & H4 & 3.8 & 6.3 & 2.5 & 5.0 & 5.2 &  \\
   & H1 & H4 & 10.0 & 12.8 & 10.1 & 9.4 & 9.9 &  \\
   & H1 & H5 & 0.9 & 1.2 & 0.7 & 1.1 & 1.1 &  \\
   & H1 & H3 & 3.9 & 5.1 & 3.8 & 1.5 & 2.0 & 4.6 \\
  \hline
  C$_2$H$_6$ & C1 & C2 & 17.3 & 28.3 & 37.5 & 24.2 & 23.1 & 34.6 \\
   & C1 & H1 & 94.0 & 122.7 & 119.8 & 92.4 & 96.2 & 124.2 \\
   & C1 & H4 & -1.2 & -2.2 & -5.3 & 0.4 & 0.6 & -4.6 \\
   & H1 & H2 & -6.5 & -10.4 & -14.1 & -3.1 & -3.3 &  \\
   & H1 & H4 & 12.4 & 16.3 & 14.7 & 11.2 & 11.8 &  \\
   & H1 & H5 & 2.8 & 3.8 & 3.5 & 2.5 & 2.6 &  \\
   & H1 & H4 & 6.0 & 8.0 & 7.2 & 5.4 & 5.7 & 8.0 \\
  \hline
  C$_2$H$_4$ & C1 & C2 & 56.4 & 73.5 & 75.7 & 64.5 & 65.4 & 67.5 \\
   & C1 & H1 & 119.2 & 155.4 & 147.7 & 116.4 & 121.1 & 156.3 \\
   & C1 & H3 & 1.2 & 0.0 & -3.3 & 2.5 & 3.0 & -2.4 \\
   & H1 & H2 & 4.4 & 4.9 & 0.9 & 7.3 & 7.4 & 2.2 \\
   & H1 & H3 & 8.6 & 11.3 & 10.4 & 7.9 & 8.0 & 11.6 \\
   & H1 & H4 & 13.0 & 18.1 & 17.0 & 11.6 & 12.4 & 19.0 \\
  \hline
  C$_2$H$_2$ & C1 & C2 & 154.5 & 176.5 & 166.5 & 155.4 & 164.1 & 184.5 \\
   & C1 & H2 & 198.7 & 254.4 & 232.1 & 186.6 & 193.4 & 242.4 \\
   & C1 & H1 & 45.6 & 53.1 & 50.1 & 42.4 & 43.9 & 53.8 \\
   & H1 & H2 & 6.2 & 9.6 & 10.8 & 4.9 & 5.0 & 10.1 \\
  \hline
  C$_6$H$_6$ & C1 & C2 & 48.4 & 63.8 & 75.1 & 54.2 & 53.8 & 55.8 \\
   & C1 & C3 & 0.7 & 0.0 & -3.7 & 1.7 & 1.9 & -2.5 \\
   & C1 & C4 & 7.9 & 8.4 & 16.8 & 7.7 & 7.9 & 10.1 \\
   & C1 & H1 & 119.8 & 155.0 & 185.1 & 117.4 & 122.0 & 158.3 \\
   & C1 & H2 & 2.5 & 2.1 & -9.8 & 4.2 & 4.7 & 1.0 \\
   & C1 & H3 & 5.5 & 7.2 & 12.9 & 5.1 & 5.5 & 7.6 \\
   & C1 & H4 & -0.3 & -0.8 & -6.1 & -0.1 & -0.2 & -1.2 \\
  \hline
  H$_2$S & H1 & H2 & -7.0 & -11.5 & -15.4 & -5.4 & -5.6 &  \\
  \hline
  HCl & H1 & Cl1 & 25.2 & 21.6 & 36.3 & 28.4 & 31.0 & 41.0 \\
  \hline
 \end{tabular}
 \end{ruledtabular}
\end{table*}

\subsection{\label{sec:moncho}Validation against existing quantum chemistry: Relativistic}

We make a comparison against a comprehensive benchmark performed by Moncho and Autschbach\cite{moncho_relativistic_2010} on a set of 45 molecules containing sixth row elements. For this study they used the Amsterdam Density Functional code (ADF) at the scalar-relativistic and spin-orbit levels of ZORA theory and with both the PBE\cite{perdew_generalized_1996} (more precisely a mix of VWN\cite{vosko_accurate_1980} and PBE) and PBE0\cite{adamo_toward_1999} functionals. As neither PBE0 nor spin-orbit effects have yet been implemented for the calculation of NMR parameters in periodic DFT calculations we compare only to their PBE scalar-relativistic ZORA calculations.

We now highlight some of the physical and numerical differences between the scalar relativistic approach used in Ref. \onlinecite{moncho_relativistic_2010} and our planewave-pseudopotential formalism. In Ref. \onlinecite{moncho_relativistic_2010} the SD term is neglected due to computational expense, however, we include it in our calculation. It is generally less than $1\%$ of the final isotropic value, though in the case of WF$_6$ it accounts for $24\%$ of the coupling.

The ADF calculations took $V$ from four-component numerical Dirac equation calculations on isolated, neutral atoms. This potential was then used to determine $\mathcal{K}$ for a ZORA Hamiltonian. Similarly, in our implementation, $\mathcal{K}$ for each species of atom is determined by the all-electron potential of a four-component calculation on an isolated, neutral atom. This was then used to construct a scalar ZORA isolated atom and so generate an ultrasoft pseudopotential.

The norm-conserving relativistic potentials were generated with the \texttt{atom} code as maintained by Jos\'e Lu\'is Martins from four-component calculations, with the all-electron partial waves taking the spin-up electron state of the valence orbitals.

By using pseudopotentials we are implicitly using the frozen core approximation. While Moncho and Autschbach do not use the frozen core approximation, Autschbach\cite{autschbach_nuclear_2000} notes that the frozen core approximation yields almost the same couplings as the respective all-electron computations as long as a sufficiently complete basis set is used. We neglect finite nucleus effects, which can be relevant for row-six elements, as noted in Section~\ref{sec:theory-j}. Finally, we also neglect solvent effects, which are included in the ADF calculations with the conductor-like screening model (COSMO) \cite{pye_implementation_1999}. However, Moncho and Autschbach find that the COSMO model does not significantly improve the median relative deviations of couplings with experiment over the gas phase calculations and suggest that explicit solvent models are necessary. We also note that the main goal of our methodology is to compute J-couplings in solid materials, where solvent effects are not relevant.

\begin{figure}
 \includegraphics[height=0.75\linewidth]{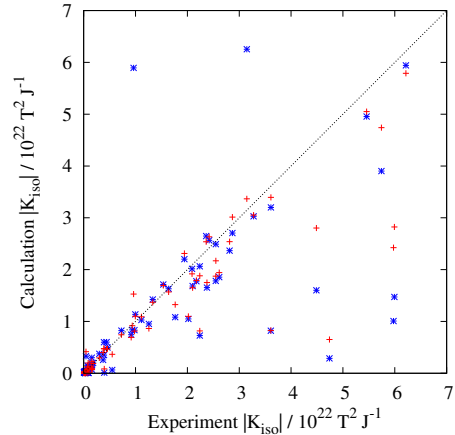}
 \caption{\label{fig:KexpVadfVcastep}Comparison of reduced isotropic J-couplings in a number of small molecules containing row six elements calculated using ADF's implementation of scalar-relativistic ZORA J-coupling (red crosses), from Ref.~\onlinecite{moncho_relativistic_2010}, and the authors' implementation in CASTEP of the same (blue stars), both using the PBE exchange-correlation functional, against experiment. Two points are excluded due to scale: \jd(Hg-Sn) in Hg-Ir-SnCl$_3$; and \js(Pt-C) in Pt(SnCl$_3$).}
\end{figure}

\begin{figure}
 \includegraphics[height=0.75\linewidth]{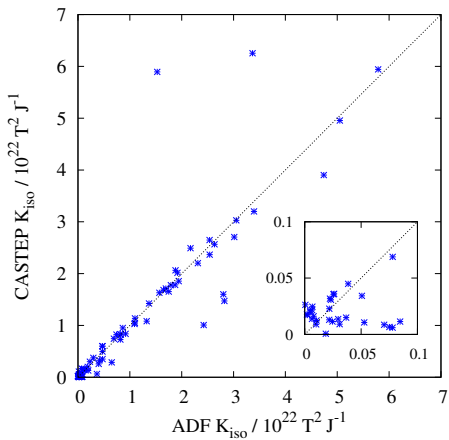}
 \caption{\label{fig:KadfVcastep}Comparison of reduced isotropic J-couplings in a number of small molecules containing row six elements calculated using ADF's implementation of scalar-relativistic ZORA J-coupling, from Ref.~\onlinecite{moncho_relativistic_2010}, against the authors' implementation in CASTEP of the same, both using the PBE exchange-correlation functional. Two points are excluded due to scale: \jd(Hg-Sn) in Hg-Ir-SnCl$_3$; and \js(Pt-C) in Pt(SnCl$_3$).}
\end{figure}

The PBE functional was used with a converged planewave basis set cut-off of 80 rydberg and a single k-point in each case. As we are implicitly using periodic boundary conditions, we must place the molecules in a supercell of sufficient size to reduce interactions between periodic images. A cubic cell size of $(15$\r{A}$)^3$ is found to be sufficient in most cases. The calculated isotropic reduced couplings for tungsten, lead, mercury, platinum and thallium are shown in Tables \ref{tab:rel_benchmark_W}, \ref{tab:rel_benchmark_Pb}, \ref{tab:rel_benchmark_Hg}, \ref{tab:rel_benchmark_Pt} and \ref{tab:rel_benchmark_Tl} respectively. Fig.~\ref{fig:KexpVadfVcastep} shows CASTEP and ADF calculations against experiment and Fig.~\ref{fig:KadfVcastep} shows the CASTEP implementation against ADF. To be consitent with Ref.~\onlinecite{moncho_relativistic_2010} in this section we report the reduced coupling constants in S.I. units (\kunits). 

Considering the full set of compounds we find that, as expected, the non-relativitistic pseudopotential calculations show large deviation from experiment, with a RMSD of 4529.2, mean absolute deviation of 1352.3 and median absolute deviation of 224.1 $\times$ \kunits. Using relativistic ultrasoft pseudopotentials and the ZORA formalism of Section \ref{sec:theory-j} significantly improves the agreement with experiment, giving a RMSD of 1215.2, mean absolute deviation of 511.6 and median absolute deviation of 87.6. These results are comparable to the equivalent results of  Moncho and Autschbach \cite{moncho_relativistic_2010} (1558.7, 517.8, 82.1). We also note that there is good agreement between the norm-conserving and ultrasoft pseudopotential results.

For tungsten containing molecules (Table \ref{tab:rel_benchmark_W}), the performance of ultrasoft pseudopotentials was equivalent to the all-electron calculations, with marginally larger RMSD and smaller mean and median absolute deviations from experiment. Lead-containing molecules (Table \ref{tab:rel_benchmark_Pb}) also gave equivalent performance for ultrasoft pseudopotentials, with marginally smaller RMSD, mean and median absolute deviations. Mercury-containing molecules (Table \ref{tab:rel_benchmark_Hg}) have a significantly smaller RMSD, dominated by the size of the \jd(Hg-Sn) coupling in IrCl(SnCl$_3$)(HgCl)(CO)(PH$_3$)$_2$, and smaller mean and median absolute deviations. Platinum-containing molecules (Table \ref{tab:rel_benchmark_Pt}) have a larger RMSD, mean and median deviations, largely due to the significant error in the \js(Pt-Pt) coupling in the charged molecules. We believe this error arises from the difficulty in treating charged systems with periodic boundary conditions and the influence of the solvation model on the all-electron couplings.

For thallium containing molecules (Table \ref{tab:rel_benchmark_Tl}), the statistics are  poor compared to experiment and show some deviations from the all-electron results. It was previously found \cite{moncho_relativistic_2010} that greater accuracy for couplings in thallium-containing molecules requires going to PBE0 and spin-orbit levels of theory. The reported couplings in molecules TlF, TlCl, TlBr and TI are taken from gas phase experiments and, except for TlI, are in good agreement with all-electron calculations. The \js(Tl-I) coupling exhibits a very large orbital current contribution, an order of magnitude larger than the total value, and a large Fermi-contact-like contribution with opposite sign. The couplings in the remaining thallium molecules are measured in solution; Tl$_4$(OCH$_3$)$_4$ was in a toluene solvent and the rest in a water solvent. Calculations presented at the PBE0 and spin-orbit level of theory in Ref.~\onlinecite{moncho_relativistic_2010} show that, in contrast to other elements, couplings in thallium containing molecules are significantly affected by the inclusion of solvent, with the mean absolute deviation going from 1163.0 to 737.1 when moving to the COSMO model. Neglect of solvation effects is hence likely to be the main source of disagreement between our results and all-electron couplings.

\begin{table*}[p!]
 \scriptsize
 \caption{\label{tab:rel_benchmark_W}Relativistic benchmark on tungsten containing molecules. Local orbital PBE ZORA calculations using ADF and experimental numbers from Ref.~\onlinecite{moncho_relativistic_2010}. Ultrasoft PBE ZORA values calculated with the implementation in CASTEP. All values are isotropic reduced J-coupling in \kunits.}
 \begin{ruledtabular}
 \begin{tabular}{lllrrrrrr}
  Structure & Site A & Site B & Exp & ADF rel & USP rel \\
  \hline
  W(CO)$_6$ & W1 & C1 & 997.0 & 1099.5 & 1138.2 \\
  \hline
  W(CCH$_3$)(CH$_2$CH$_3$)$_3$ & W1 & H10 & 20.7 & -24.6 & -11.4 \\
  \hline
  W(CO)$_5$PI$_3$ & W1 & P1 & 1639.0 & 1568.7 & 1633.1 \\
  \hline
  $\eta$-(C$_5$H$_5$)W(CO)$_3$H & W1 & H6 & 72.4 & 127.4 & 148.2 \\
  \hline
  W(CO)$_5$PF$_3$ & W1 & P1 & 2362.9 & 2537.3 & 2651.9 \\
  \hline
  W(CO)$_5$PCl$_3$ & W1 & P1 & 2090.0 & 1918.0 & 2018.1 \\
  \hline
  WF$_6$ & W1 & F1 & 85.4 & -198.5 & -131.4 \\
  \hline
  \multicolumn{4}{r}{Root mean square deviation} & 114.2 & 129.1 \\
  \multicolumn{4}{r}{Mean absolute deviation} & 98.7 & 91.3 \\
  \multicolumn{4}{r}{Median absolute deviation} & 102.5 & 71.9 \\
 \end{tabular}
 \end{ruledtabular}
\end{table*}

\begin{table*}[p!]
 \scriptsize
 \caption{Relativistic benchmark on lead containing molecules. Local orbital PBE ZORA calculations using ADF and experimental numbers from Ref.~\onlinecite{moncho_relativistic_2010}. Ultrasoft PBE ZORA values calculated with the implementation in CASTEP. All values are isotropic reduced J-coupling in \kunits.}
 \begin{ruledtabular}
 \begin{tabular}{lllrrrrrr}
  Structure & Site A & Site B & Exp & ADF rel & USP rel \\
  \hline
  Pb(CH$_3$)$_3$CF$_3$ & Pb1 & F1 & 102.0 & 17.1 & 109.8 \\
   & Pb1 & H1 & 28.7 & -4.9 & 21.0 \\
  \hline
  PbCl$_4$ & Pb1 & Cl1 & 2868.3 & -3014.1 & -2709.0 \\
  \hline
  Pb(CH$_3$)$_2$(CF$_3$)$_2$ & Pb1 & F1 & 160.7 & 70.4 & 169.8 \\
   & Pb1 & H1 & 33.9 & -6.8 & 24.6 \\
  \hline
  PbH$_4$ & Pb1 & H1 & 1115.0 & 1090.7 & 1030.0 \\
  \hline
  Pb(CH$_3$)$_2$H$_2$ & Pb1 & H1 & 979.4 & 814.0 & 832.6 \\
   & Pb1 & H3 & -30.3 & -10.1 & 8.9 \\
  \hline
  Pb$_2$(CH$_3$)$_6$ & Pb1 & C1 & 44.4 & -416.5 & -328.1 \\
   & Pb1 & C6 & 146.0 & 121.7 & 152.7 \\
   & Pb1 & H1 & -16.8 & 5.0 & 19.9 \\
   & Pb1 & H16 & 9.1 & 10.5 & 12.4 \\
  \hline
  Pb(CH$_3$)$_3$H & Pb1 & H1 & 915.9 & 692.5 & 744.9 \\
   & Pb1 & H2 & -27.1 & -6.4 & 13.8 \\
  \hline
  Pb(CH$_3$)$_4$ & Pb1 & C1 & 396.7 & -84.6 & -11.5 \\
   & Pb1 & H1 & 24.5 & -2.3 & 17.2 \\
  \hline
  \multicolumn{4}{r}{Root mean square deviation} & 148.7 & 139.9 \\
  \multicolumn{4}{r}{Mean absolute deviation} & 98.1 & 82.5 \\
  \multicolumn{4}{r}{Median absolute deviation} & 25.7 & 11.3 \\
 \end{tabular}
 \end{ruledtabular}
 \label{tab:rel_benchmark_Pb}
\end{table*}

\begin{table*}[p!]
 \scriptsize
 \caption{\label{tab:rel_benchmark_Hg}Relativistic benchmark on mercury containing molecules. Local orbital PBE ZORA calculations using ADF and experimental numbers from Ref.~\onlinecite{moncho_relativistic_2010}. Ultrasoft PBE ZORA values calculated with the implementation in CASTEP. All values are isotropic reduced J-coupling in \kunits.}
 \begin{ruledtabular}
 \begin{tabular}{lllrrrrrr}
  Structure & Site A & Site B & Exp & ADF rel & USP rel \\
  \hline
  Hg(CH$_3$)Cl & Hg1 & C1 & 2614.5 & 1943.0 & 1857.1 \\
   & Hg1 & H1 & 93.6 & -77.9 & 6.0 \\
  \hline
  $[$Hg(CN)$_4]^{2-}$ & Hg1 & C1 & 2814.2 & 2540.4 & 2368.4 \\
  \hline
  Hg(CCCl)$_2$ & Hg1 & C1 & 5456.6 & 5049.0 & 4961.6 \\
   & Hg1 & C3 & 1535.8 & 1694.3 & 1714.7 \\
  \hline
  Hg(C$_6$H$_5$)$_2$ & Hg1 & C1 & 2174.6 & 1791.5 & 1778.7 \\
   & Hg1 & C2 & 157.0 & 228.9 & 301.7 \\
   & Hg1 & C3 & 183.7 & 188.8 & 199.7 \\
   & Hg1 & C4 & 32.0 & -29.7 & -13.8 \\
  \hline
  Hg(CH$_3$)CCH & Hg1 & C1 & 2549.2 & 2169.9 & 2493.6 \\
   & Hg1 & C2 & 2101.5 & 1650.0 & 1685.2 \\
   & Hg1 & C3 & 728.0 & 743.6 & 823.3 \\
  \hline
  Hg(CH$_3$)I & Hg1 & C1 & 2378.0 & 1751.2 & 1655.5 \\
   & Hg1 & H1 & 84.6 & -70.5 & 8.6 \\
  \hline
  IrCl(SnCl$_3$)(HgCl) $\ldots$ & Hg1 & Sn1 & -50838.5 & 62548.2 & 50861.4 \\
  $\ldots$ (CO)(PH$_3$)$_2$ & Hg1 & P1 & 375.5 & -399.1 & -257.2 \\
  \hline
  Hg(CH$_3$)(CF$_3$) & Hg1 & F1 & 459.0 & 471.7 & 496.3 \\
   & Hg1 & H1 & 64.0 & -52.8 & 10.7 \\
  \hline
  Hg(CH$_3$)$_2$ & Hg1 & C1 & 1258.2 & 863.5 & 953.0 \\
   & Hg1 & H1 & 46.4 & -36.7 & 14.7 \\
  \hline
  Hg(CH$_3$)Br & Hg1 & C1 & 2546.7 & 1873.8 & 1785.1 \\
   & Hg1 & H1 & 90.5 & -75.1 & 6.6 \\
  \hline
  Hg(CN)$_2$ & Hg1 & C1 & 5741.7 & 4740.7 & 3907.1 \\
  \hline
  \multicolumn{4}{r}{Root mean square deviation} & 2470.2 & 511.4 \\
  \multicolumn{4}{r}{Mean absolute deviation} & 753.4 & 311.1 \\
  \multicolumn{4}{r}{Median absolute deviation} & 158.5 & 118.3 \\
 \end{tabular}
 \end{ruledtabular}
\end{table*}

\begin{table*}[p!]
 \scriptsize
 \caption{\label{tab:rel_benchmark_Pt}Relativistic benchmark on platinum containing molecules. Local orbital PBE ZORA calculations using ADF and experimental numbers from Ref.~\onlinecite{moncho_relativistic_2010}. Ultrasoft PBE ZORA values calculated with the implementation in CASTEP. All values are isotropic reduced J-coupling in \kunits.}
 \begin{ruledtabular}
 \begin{tabular}{lllrrrrrr}
  Structure & Site A & Site B & Exp & ADF rel & USP rel \\
  \hline
  cis-PtH$_2$(P(CH$_3$)$_3$)$_2$ & Pt1 & H1 & 392.1 & 474.2 & 599.3 \\
   & Pt1 & P1 & 1765.2 & 1324.9 & 1083.9 \\
  \hline
  $[$Pt(CO)$_3]_2^{2+}$ & Pt1 & C1 & 2420.2 & 2632.3 & 2576.4 \\
   & Pt1 & C3 & 1943.6 & 2312.0 & 2204.5 \\
   & Pt1 & C4 & 39.7 & -38.5 & -44.9 \\
   & Pt1 & C6 & 302.7 & 296.8 & 375.2 \\
   & Pt1 & Pt2 & 962.9 & 1527.9 & 5893.1 \\
  \hline
  Pt(PF$_3$)$_4$ & Pt1 & P1 & 6215.0 & 5789.7 & 5949.6 \\
  \hline
  Pt(P(CH$_3$)$_3$)$_4$ & Pt1 & P1 & 3610.5 & 3393.7 & 3205.5 \\
  \hline
  trans-PtI$_2$(NH$_2$CH$_3$)$_2$ & Pt1 & C1 & 22.8 & -6.5 & 21.7 \\
   & Pt1 & H1 & 23.3 & -21.7 & -12.9 \\
  \hline
  cis-PtCl$_2$(P(CH$_3$)$_3$)$_2$ & Pt1 & P1 & 3276.2 & 3053.5 & 3033.2 \\
  \hline
  cis-PtI$_2$(NH$_2$CH$_3$)$_2$ & Pt1 & C1 & 31.9 & -18.7 & -0.6 \\
   & Pt1 & H1 & 26.3 & -23.3 & -30.8 \\
  \hline
  Pt(SnCl$_3$)(CH$_2$C$\ldots$ & Pt1 & Sn1 & -28025.3 & 23956.7 & 26441.7 \\
  $\ldots$ (CH$_3$)CH$_2$)(CH$_2$H$_4$) & Pt1 & C4 & 50.1 & -50.7 & -34.2 \\
   & Pt1 & C2 & 65.2 & -0.6 & 26.3 \\
   & Pt1 & C3 & 160.8 & 78.0 & 68.9 \\
   & Pt1 & C1 & 157.7 & 95.3 & 151.1 \\
   & Pt1 & C6 & 119.8 & 104.5 & 141.2 \\
   & Pt1 & H1 & 19.5 & -22.4 & -31.8 \\
   & Pt1 & H2 & 16.8 & -7.9 & -16.2 \\
   & Pt1 & H3 & 23.7 & -26.1 & -36.1 \\
   & Pt1 & H4 & 22.5 & -25.5 & -35.6 \\
   & Pt1 & H5 & 6.1 & -2.0 & -17.8 \\
   & Pt1 & H6 & 22.5 & 31.0 & 9.1 \\
  \hline
  trans-PtH$_2$(P(CH$_3$)$_3$)$_2$ & Pt1 & H1 & 392.1 & 474.2 & 351.1 \\
  \hline
  trans-PtCl$_2$(P(CH$_3$)$_3$)$_2$ & Pt1 & P1 & 2239.7 & 1880.4 & 2067.0 \\
  \hline
  $[$Pt(CN)$_5]_2^{4-}$ & Pt1 & C1 & 1331.7 & 1369.1 & 1426.7 \\
   & Pt1 & C5 & 937.3 & 918.8 & 839.2 \\
   & Pt1 & C6 & 30.0 & -21.8 & -22.9 \\
   & Pt1 & C10 & 441.4 & 468.9 & 601.9 \\
   & Pt1 & Pt2 & 3146.0 & 3367.4 & 6254.0 \\
  \hline
  \multicolumn{4}{r}{Root mean square deviation} & 733.1 & 1065.4 \\
  \multicolumn{4}{r}{Mean absolute deviation} & 231.9 & 387.2 \\
  \multicolumn{4}{r}{Median absolute deviation} & 27.5 & 41.0 \\
 \end{tabular}
 \end{ruledtabular}
\end{table*}

\begin{table*}[p!]
 \scriptsize
 \caption{\label{tab:rel_benchmark_Tl}Relativistic benchmark on thallium containing molecules. Local orbital PBE ZORA calculations using ADF and experimental numbers from Ref.~\onlinecite{moncho_relativistic_2010}. Ultrasoft PBE ZORA values calculated with the implementation in CASTEP. All values are isotropic reduced J-coupling in \kunits.}
 \begin{ruledtabular}
 \begin{tabular}{lllrrrrrr}
  Structure & Site A & Site B & Exp & ADF rel & USP rel \\
  \hline
  Tl(CN)$_3$ & Tl1 & C1 & 4488.5 & 2802.2 & 1601.2 \\
  \hline
  TlBr & Tl1 & Br1 & -3610.0 & -825.9 & -821.1 \\
  \hline
  Tl(CN)Cl$_2$ & Tl1 & C1 & 5975.5 & 2424.9 & 1008.2 \\
  \hline
  Tl(CN)$_2$Cl & Tl1 & C1 & 5991.8 & 2822.9 & 1475.5 \\
  \hline
  Tl$_4$(OCH$_3$)$_4$ & Tl1 & Tl1 & 549.8 & -365.3 & -63.2 \\
  \hline
  TlCl & Tl1 & Cl1 & -2240.0 & -817.4 & -728.6 \\
  \hline
  TlF & Tl1 & F1 & -2020.0 & -1096.7 & -1050.0 \\
  \hline
  TlI & Tl1 & I1 & -4740.0 & -652.2 & -286.8 \\
  \hline
  \multicolumn{4}{r}{Root mean square deviation} & 2570.6 & 3249.6 \\
  \multicolumn{4}{r}{Mean absolute deviation} & 2226.0 & 2822.6 \\
  \multicolumn{4}{r}{Median absolute deviation} & 2235.2 & 2838.1 \\
 \end{tabular}
 \end{ruledtabular}
\end{table*}

\subsection{\label{sec:silver}Example: Prediction of J-coupling in silver triphenylphosphine systems}

\begin{figure}
 \includegraphics[width=0.75\linewidth]{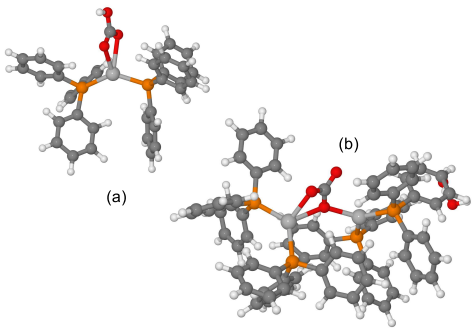}
 \caption{\label{fig:gb1054structure}Molecular structure of (a) (Ph$_3$P)$_2$Ag(O$_2$COH) and (b) [\{(Ph$_3$P)$_2$Ag\}$_2$(CO$_3$)]$\cdot$2H$_2$O. J-coupling parameters are given in Table \ref{tab:Ag-comparison}} 
\end{figure}

Two molecular crystals of interest are (Ph$_3$P)$_2$Ag(O$_2$COH) and [\{(Ph$_3$P)$_2$Ag\}$_2$(CO$_3$)]$\cdot$2H$_2$O, shown in Figure \ref{fig:gb1054structure}, part of a family of compounds incorporating phosphorous atoms bonded via a silver atom, which is in turn bonded to a varying complex. These are useful model systems for studying J-coupling between a heavy atom and a lighter atom. The J-couplings were experimentally measured with $^{31}$P CPMAS NMR by Bowmaker et al.\cite{bowmaker_solution_2011} and are quoted for both $^{107}$Ag and $^{109}$Ag when the two splittings were distinguishable. When converted to reduced couplings this gives a range of values, which could be considered as an indication of the error in the measurements.

We have calculated J-couplings using both non-relativistic theory and ZORA scalar-relativistic theory with ultrasoft pseudopotentials for both structures with a 80 rydberg kinetic energy cutoff and a single k-point with primitive cells containing 148 atoms and 296 atoms respectively. Table \ref{tab:Ag-comparison} shows the calculated couplings. Two sets of calculations for each structure are given: one for which only the hydrogen atomic positions have been optimized; and the other for which all atomic positions have been optimized, both using ultrasoft pseudopotentials with an energy cut-off of 50 rydberg. In both cases the lattice vectors were kept fixed to their experimental values. To illustrate the cost of treating such a system, each calculation at a single perturbing atom took approximately 4,500 seconds for (Ph$_3$P)$_2$Ag(O$_2$COH) and 25,000 seconds for [\{(Ph$_3$P)$_2$Ag\}$_2$(CO$_3$)]$\cdot$2H$_2$O on a cluster containing four dual-quad core Intel Xeon E5620 processors, giving a total of 32 cores.
 
When a relativistic pseudopotential is used on the silver atom the $^2$J($^{31}$P-$^{31}$P) couplings are in good agreement with experiment. The difference in $^2$J($^{31}$P-$^{31}$P) coupling between the relativistic and non-relativistic cases demonstrate indirect effects in the J-coupling caused by changes in the ground state electronic structure due to use of a relativistic pseudopotential on the silver atom, as noted in previous studies involving two-bond couplings through a heavy ion\cite{enevoldsen_relativistic_2000}. The $^1$J($^{107/109}$Ag-$^{31}$P) couplings are in reasonable agreement with experiment when ZORA-level corrections to the operators are applied. The computed couplings agree better with the lower end of the quoted experimental ranges. The $^1$J($^{109}$Ag-$^{31}$P) couplings demonstrate direct effects due to use of ZORA operators when performing the perturbation as well as indirect ground state electronic structure effects. Performing a full optimization of all the atomic positions gives only a small improvement in the calculated couplings. For [\{(Ph$_3$P)$_2$Ag\}$_2$(CO$_3$)]$\cdot$2H$_2$O, the calculations give a spread of $^1$J(Ag-P) couplings corresponding to the four crystallographically distinct phosporous sites - experimentally only a single coupling was determined, presumably an average value.

\begin{table*}[p!]
\scriptsize
\caption{\label{tab:Ag-comparison}Calculated and experimental\cite{bowmaker_solution_2011} isotropic reduced coupling constants (K), in \kunits, for (Ph$_3$P)$_2$Ag(O$_2$COH) and [\{(Ph$_3$P)$_2$Ag\}$_2$(CO$_3$)]$\cdot$2H$_2$O. Their structures are shown in Figure \ref{fig:gb1054structure}. For $K_{\textrm{Calc}}^{\textit{ZORA}}$ silver atoms were treated with relativistic pseudopotentials and ZORA modified operators.}

\begin{ruledtabular}
\begin{tabular}{rllllll}

Structure (optimization) & Site A & Site B & $K_{\textrm{Calc}}^{\textit{NRel}}$ & $K_{\textrm{Calc}}^{\textit{ZORA}}$ & $K^{\textrm{Ag109}}_{\textrm{Exp}}$ & $K^{\textrm{Ag107}}_\textrm{Exp}$\\

\hline

(Ph$_3$P)$_2$Ag(O$_2$COH) & Ag     & P 1 & 1,481 & 1,952 & \multirow{2}{*}{2,059} & \multirow{2}{*}{2,398} \\
All ions                  & Ag     & P 2 & 1,492 & 1,958 & & \\
                          & P 1    & P 2 & 53.0  & 70.6  & 74.5 & 74.5 \\

(Ph$_3$P)$_2$Ag(O$_2$COH) & Ag     & P 1 & 1,462 & 1,932 & \multirow{2}{*}{2,059} & \multirow{2}{*}{2,398} \\
Only hydrogen             & Ag     & P 2 & 1,462 & 1,925 & & \\
                          & P 1    & P 2 & 52.0 & 69.2  & 74.54 & 74.54 \\

\hline
[\{(Ph$_3$P)$_2$Ag\}$_2$(CO$_3$)]$\cdot$2H$_2$O     & Ag 1   & P 1 & 1,259 & 1,644 & \multirow{8}{*}{2,125*} & \multirow{8}{*}{1,848*} \\
All ions                                            & Ag 1   & P 2 & 1,398 & 1,849 &   & \\
                                                    & Ag 2   & P 3 & 1,470 & 1,919 &   & \\
                                                    & Ag 2   & P 4 & 1,229 & 1,583 &   & \\

[\{(Ph$_3$P)$_2$Ag\}$_2$(CO$_3$)]$\cdot$2H$_2$O     & Ag 1   & P 1 & 1,259 & 1,650 &   & \\
Only hydrogen                                       & Ag 1   & P 2 & 1,350 & 1,786 &   & \\
                                                    & Ag 2   & P 3 & 1,433 & 1,887 &   & \\
                                                    & Ag 2   & P 4 & 1,145 & 1,467 &   & \\

\end{tabular}
* Ag$^{107}$ and Ag$^{109}$ couplings were not resolved separately in experiment.
\end{ruledtabular}
\end{table*}

\section{\label{sec:conclusion}Conclusions}

We have presented a method for the calculation of J-coupling tensors for systems containing heavy elements using state-of-the-art ultrasoft pseudopotentials which gives accuracy comparable to existing quantum chemistry methods at the same level of theory. The use of pseudopotentials allows fewer electrons to be explicitly treated, and we have shown that the treatment of relativity using the ZORA approach allows cheap corrections to be made using PAW. This allows calculations of J-coupling tensors between heavy ions in periodic and molecular systems containing hundreds of atoms.

\section{\label{sec:si}Supplementary materials}

Compressed archives of calculation input and output files, including structures and magnetic resonance tensors, are available online for the non-relativistic benchmark\cite{si_nrel}, relativistic benchmark\cite{si_rel} and silver-containing molecular crystals\cite{si_ag}.

\begin{acknowledgments}
The authors acknowledge financial support from EPSRC (TFGG, JRY) and the Royal Society (JRY).
\end{acknowledgments}

\end{document}